\documentclass[10.5pt,a4paper,aps,pra,preprintnumbers,amsmath,amssymb,twocolumn,longbibliography,superscriptaddress]{revtex4-2}
\usepackage{mathtools}
\usepackage{amsmath}
\usepackage{graphicx}
\usepackage{bbm}
\usepackage{dcolumn}
\usepackage{units}
\usepackage{bm}
\usepackage{hyperref}
\usepackage{url}
\usepackage{xfrac}
\usepackage{amssymb}
\usepackage{bbold}
\usepackage[mathlines]{lineno}

\usepackage{graphicx}
\usepackage{dcolumn}
\usepackage{bm}
\usepackage{xcolor}
\DeclareUnicodeCharacter{202F}{\,}
\newcommand{\ket}[1]{|#1\rangle}

\begin{document}

\title{Quantum Algorithms for Inverse Participation Ratio Estimation in multi-qubit and multi-qudit systems}

\author{Yingjian Liu}
\affiliation{ICFO-Institut de Ciencies Fotoniques, The Barcelona Institute of Science and Technology, 08860 Castelldefels (Barcelona), Spain}

\author{Piotr Sierant}
\affiliation{ICFO-Institut de Ciencies Fotoniques, The Barcelona Institute of Science and Technology, 08860 Castelldefels (Barcelona), Spain}

\author{Paolo Stornati}
\affiliation{ICFO-Institut de Ciencies Fotoniques, The Barcelona Institute of Science and Technology, 08860 Castelldefels (Barcelona), Spain}

\author{Maciej Lewenstein}
\affiliation{ICFO-Institut de Ciencies Fotoniques, The Barcelona Institute of Science and Technology, 08860 Castelldefels (Barcelona), Spain} 
\affiliation{ICREA, Passeig Lluis Companys 23, 08010 Barcelona, Spain}

\author{Marcin Płodzień}
\affiliation{ICFO-Institut de Ciencies Fotoniques, The Barcelona Institute of Science and Technology, 08860 Castelldefels (Barcelona), Spain}

\date{\today}

\begin{abstract}
Inverse Participation Ratios (IPRs) and the related Participation Entropies quantify the spread of a quantum state over a selected basis of the Hilbert space, offering insights into the equilibrium and non-equilibrium properties of the system. In this work, we propose three quantum algorithms to estimate IPRs on multi-qubit and multi-qudit quantum devices. The first algorithm allows for the estimation of IPRs in the computational basis by single-qubit measurements, while the second one enables measurement of IPR in the eigenbasis of a selected Hamiltonian, without the knowledge about the eigenstates of the system. Next, we provide an algorithm for IPR in the computational basis for a multi-qudit system. We discuss resources required by the algorithms and benchmark them by investigating the one-axis twisting protocol, the thermalization in a deformed PXP model, and the ground state of a spin-$1$ AKLT (Affleck-Kennedy-Lieb-Tasaki) chain in a transverse field. 
\end{abstract}

\maketitle


\section{Introduction}

Understanding equilibrium~\cite{Amico08, Eisert10area, Schollwock11, Alhambra23, turkeshi2024hilbert, sierant2025many} and non-equilibrium \cite{eisert2015quantum, Dalessio16, Bertini21transport, Abanin19MBLrev, Fisher2023, sierant2025many} properties of quantum many-body systems is a significant challenge in contemporary physics~\cite{anderson1972more}. While investigations of many-body systems on classical computers are hindered by the exponential growth of the Hilbert space dimension with the system size, the recent experimental breakthroughs in realizing and controlling synthetic matter platforms fulfill the vision of quantum simulation~\cite{Lewenstein07, Gross17, Browaeys20rev, Fraxanet23} proposed by R. Feynman~\cite{feynman1982}. Moreover, quantum computing has already reached a point in which non-trivial computational tasks may be performed in gate-based settings on quantum processors \cite{Arute19, Wu21, Kim2023} despite operating in the presence of noise and errors and still belonging to the Noisy Intermediate-Scale Quantum (NISQ) era~\cite{NISQ}.
 
The research of quantum algorithms, whose paradigmatic examples include algorithms for integer factorization~\cite{Shor94}, unstructured database search~\cite{Grover96} or quantum phase estimation algorithm ~\cite{kitaev1995quantum}, has been intensified due to the advancements of quantum processors during the NISQ era, as reviewed in~\cite{Montanaro16, Bharti22rev}. Examples include quantum algorithmic solutions for quantum dynamics~\cite{miessen2021quantum, miessen2023quantum}  allowing for simulation of time evolution of many-body systems~\cite{zalka1998efficient, cleve2019efficient}, finding the energy spectrum of a static Hamiltonian~\cite{aspuru2005simulated, Wang_2008}, or approximating thermal states~\cite{PhysRevA_108_022612, Sampling, wang2021variational}.
These explorations not only open the door for discoveries in large-scale quantum many-body systems~\cite{jones2019variational,  MBLsimulatiors, chen2022error, liu2023probing, andrade2024observing} but also provide ways of benchmarking the quantum hardware \cite{mccaskey2019quantum}. Variational quantum algorithms that leverage classical optimization techniques to train parameterized quantum circuits~\cite{vqe, qaoa, Cerezo21rev, dawid2023modern} constitute another group of methods aimed at addressing the constraints of a limited number of qubits and noise characteristic for quantum processors from the NISQ era.

Among various types of quantum algorithms, approaches aimed at quantifying the properties of the quantum state are of pivotal importance for building, calibrating, and controlling quantum systems~\cite{bialas2024renyi, heightman2024solving}. Quantum state tomography, i.e., a complete reconstruction of a quantum state~\cite{Vogel89, Cramer10tomography} has limited applications in many-body systems due to the exponential growth of the Hilbert space dimension with qubit number. This led to the development of approaches such as shadow tomography~\cite{Huang2020} or randomized measurement toolbox~\cite{Elben23}  aimed at quantifying specific properties of quantum states including higher order correlation functions~\cite{pedernales2014efficient}, entanglement entropies \cite{entropy2017, Dalmonte18, Brydges19, wang2023quantum}, out-of-time-order correlations~\cite{garttner2017measuring, mi2021information} or stabilizer Renyi entropies \cite{Haug23efficient} which quantify the non-Clifford resources required to prepare the state~\cite{Leone22sre}. 
 
In this work, we focus on the inverse participation ratios (IPRs) and the related participation entropies, which quantify the spread of state $|\psi \rangle$ in a selected basis $\mathcal{B}$ of the Hilbert space of quantum many-body system. We propose quantum algorithms to estimate IPRs in the computational basis and in the eigenbasis of a selected Hamiltonian. The introduced algorithms are benchmarked with exact numerical computations.
This paper is organized as follows. In Sec.~\ref{sec:II}, we introduce the notions of IPRs and participation entropies, commenting on their relevance for quantum many-body systems. Next, in Sec.~\ref{sec:III}, we propose quantum algorithms to estimate IPR in the computational basis and in the eigenbasis of a Hamiltonian for qubits and qudits. Relevant examples, numerical results, and simulations on quantum processors are presented in Sec.~\ref{sec:IV}. In Sec.~\ref{sec:V}, we conclude and discuss the utility of the introduced algorithms for near-term quantum computing.
 
\section{Inverse Participation Ratios}\label{sec:II}
 
Let us consider the arbitrary many-body hermitian operator $\hat{A}$ and the complete many-body basis ${\cal B}_A = \{|i\rangle\}$ in which operator $\hat{A}$ is diagonal. Any pure quantum state $|\psi\rangle$ can be decomposed as $|\psi\rangle = \sum_{i} c_{i} |i\rangle$, where $c_{i} = \langle i | \psi \rangle$. To quantify properties of $|\psi\rangle$, we consider the IPRs defined as
\begin{equation}
    I^{A}_q = \sum_{i=0}^{{\cal N}-1} |c_{i}|^{2q}, \quad q=2,3,4 \cdots,
    \label{eq:IQ}
\end{equation}
where ${\cal N} = d^L$ is the dimension of the Hilbert space, $d$ is local Hilbert space size, and the integer $q \geq 2$ is referred to as the R\'{e}nyi index. The IPR $I^{A}_q$ takes values in the range $[{\cal N}^{1-q}, 1]$. The lower bound corresponds to the case when $|\psi\rangle$ uniformly populates all the basis states, namely, $|c_{i}|^2 ={\cal N}^{-1}$. The upper bound is admitted when $|\psi\rangle$ is fully localized on a single basis state $|j\rangle$, i.e., $c_{i} =\delta_{i,j}$. The IPRs have been one of the main tools in assessing the localization properties of single particle wave functions across the Anderson localization transition~\cite{Thouless74, Kramer93, Evers08, Rodriguez09, Rodriguez10}, including recent investigations of the Anderson model on hierarchical graphs~\cite{Tikhonov16, Tikhonov16b, Tikhonov19, Pino20, vanoni2023renormalization, Garcia-Mata22, sierant2023universality, vanoni2024analysis}.

For a system of $L$ qubits, $d=2$,  the Hilbert space dimension is ${\cal N}=2^L$,  which implies that $I^{A}_q$ may be exponentially small in the system size $L$, motivating introduction of the participation entropies $S_q$, defined as R\'{e}nyi entropies of the probability distribution $p_i = |c_{i}|^2$,  
\begin{equation}
    S_q = \dfrac{1}{1-q} \log_2 I^{A}_q, \quad q=2,3,4 \cdots.
    \label{eq:SQ}
\end{equation}
which constitute a related measure of the spread of $|\psi\rangle$ in the basis ${\cal B}_A$. The system size dependence of the participation entropy can be parameterized, in a sufficiently narrow interval of system sizes, as $S_q = D_q \log_2 {\cal N}  + c_q$ where $D_q$ is a fractal dimension~\cite{halsey1986fractal},  $c_q$ is a sub-leading term. If the analyzed state $\ket{ \psi }$ is localized on a fixed number of basis states, the participation entropy $S_q$ is independent of $L$, resulting in a vanishing fractal dimension $D_q=0$. In contrast, a multi-qubit state uniformly delocalized over the basis ${\cal B}_A$, $|c_{i}|^2 ={\cal N} ^{-1}$, is characterized by $D_q=1$. Similarly, Haar-random states, obtained as $\ket{\psi} = U \ket{\psi_0}$, where $\ket{\psi_0}$ is a fixed state and $U$ is a matrix drawn with Haar measure from the unitary group on $L$ qubits, are fully delocalized in the Hilbert space, with the fractal dimension $D_q=1$ (and a sub-leading term $c_q<0$)~\cite{Backer19}. Multifractality~\cite{Stanley88,dziurawiec2023unraveling} is the intermediate case between the delocalization ($D_q=1$) and localization ($D_q=0$) when the fractal dimension $0<D_q<1$ depends non-trivially on the R\'enyi index $q$. We note that quantum wavelet transform was proposed in Ref.~\cite{Garcia09} as means of extracting the value of the multifractal dimensions $D_q$.

The participation entropies of ground states of quantum many-body systems have been employed to distinguish between various quantum phases~\cite{Stephan09, Stephan10, Stephan14, Luitz14universal, Luitz14improving, Pino17, Lindinger19, Pausch21,PhysRevB.109.064302, Lozano21}. Moreover, participation entropies can be used as an ergodicity-breaking measure in quantum many-body systems as proposed in~\cite{de2013ergodicity}. Indeed, while the properties eigenstates of thermalizing~\cite{ Srednicki94, Deutsch91, Rigol08, pappalardi, fava2023designs} many-body systems may be modeled by the fully delocalized random Haar states, ergodicity breaking due to many-body localization~\cite{Nandkishore15, Alet18, Abanin19MBLrev, sierant2025many} is manifested by the multifractality of many-body states~\cite{Mace19Multifractal}. Similarly, measurement-induced phase transitions in unitary dynamics of random circuits~\cite{Fisher2023} interspersed with local measurements can be investigated through the lens of participation entropies~\cite{sierant2022universal, Sierant22measurement}. The IPRs and participation entropies can also be used to probe time dynamics of quantum circuits~\cite{Turkeshi24hilbert} and are related to the relative entropies of coherence~\cite{Baumgratz} important for the resource theory of quantum coherence~\cite{Chitambar19}, and stabilizer Renyi entropies~\cite{Turkeshi23flatness}. Finally, the IPR $I_2$ coincides with a \textit{collision probability} describing the anticoncentration properties of the many-body wave function, relevant for the formal arguments of classical hardness of the sampling problems~\cite{aaronson2011computational, Bremner16, Bouland19, Oszmaniec22}.

The broad relevance of the IPRs and participation entropies to quantum many-body systems motivates us to consider quantum algorithms for their measurement.  For concreteness, since IPRs and participation entropies are functionally dependent, c.f. \eqref{eq:SQ}, we focus on measuring the IPRs~\eqref{eq:IQ}.

\section{Quantum algorithms for IPR estimation} \label{sec:III}
In this section, we introduce the primary algorithm developed within this paper. Subsection \ref{sec:IIIA} details the algorithm for computing the Inverse Participation Ratio in the computational basis. Subsequently, in Subsection \ref{sec:IIIB}, we extend this algorithm to calculate the IPR in the eigenbasis of a chosen Hamiltonian for qubits. Finally, in Subsection \ref{sec:IIIC}, we generalize the computation of the IPR in the computational basis for qudits.

\subsection{IPR in the computational basis for qubits}\label{sec:IIIA}

We consider a multi-qubit state $\ket{\psi}$ and fix the basis of interest as the computational basis, i.e. the eigenbasis of Pauli-Z operators,   $\mathcal B_Z = \{ \ket{\sigma_i} \}$, where $\sigma_i=0,1$.  A naive experimental procedure for measuring IPRs in the computational basis could consist of performing the measurements of the $\hat{Z}$ operators, recording the resulting bitstrings associated with the basis states $\ket{\sigma_i}$, estimating the probabilities $|c_i|^2 = | \langle \sigma_i  \ket{\psi} |^2$ and calculating the IPRs using their definition Eq.\eqref{eq:IQ}. While the procedure of such a sampling of the state is experimentally realized on quantum processors~\cite{Arute19} and forms a basis of the cross-entropy benchmarking~\cite{Boixo2018}, it requires estimation of exponentially many probabilities $|c_i|^2$ associated with each of the states of $\mathcal B_Z$.

\begin{figure}[t!]
	\centering
	\includegraphics[width=0.45\textwidth]{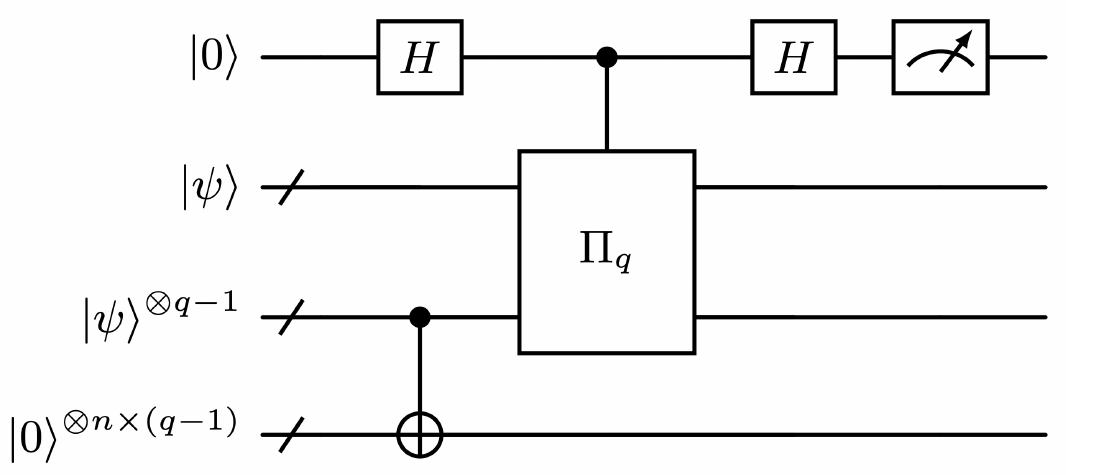} 
	\caption{Quantum circuit for estimating IPR $I^Z_q$ of R\'{e}nyi index $q\geq 2$ in the computational basis $\mathcal B_Z$. The quantum circuit comprises $n\times (q-1)$ CNOT gates and a controlled-$\Pi_q$ gate. The value of $I^Z_q$ is determined as an expectation value of the Pauli-Z operator on the ancillary qubit.}
	\label{fig:circuit1}
\end{figure}

In the following, we propose a quantum algorithm that allows the measurement of IPR in a computational basis, denoted as $I^{Z}_q$, as an expectation value of a single-qubit measurement.
 
The algorithm requires $q$ copies of the state $|\psi\rangle = \sum_{i} c_i \ket{i}$ (where the index $i$ runs over the states of the computational basis $\mathcal B_Z$) and additional $n$-qubit quantum registers, so that the input state reads

\begin{align}\label{Eq:cnot}
    |\Psi_{0}\rangle = &  |0\rangle\otimes |\psi\rangle \otimes |\psi\rangle^{\otimes q-1} \otimes |0\rangle^{ \otimes n (q-1)}, \notag\\
    = & \sum_{i_0,\cdots, i_{q-1}}c_{i_{1}} \cdots c_{i_{q-1}} |0\rangle \otimes | i_0,\cdots, i_{q-1}\rangle \otimes  |0\rangle^{ \otimes n (q-1)}.
\end{align}
The algorithm generates entanglement between the $(q-1)$ copies of $|\psi\rangle$ and the $n$-qubit quantum registers using the CNOT gates
\begin{align}\label{Eq4}
     U^{\otimes n (q-1)}_{\text{CNOT}} |\Psi_{0}\rangle =&\sum_{i_0,\cdots, i_{q-1}} c_{i_{0}} \cdots c_{i_{q-1}} |0\rangle \otimes | i_0 \rangle \notag\\
     &\otimes | i_1,\cdots, i_{q-1}\rangle \otimes | i_1,\cdots, i_{q-1}\rangle.
\end{align}
Subsequently, one ancillary qubit controls a cyclic permutation, denoted as a controlled-$\Pi_q$ gate in Figure~\ref{fig:circuit1}. $\Pi_q$ acts on the $q$ systems as
\begin{equation}
    \Pi_q   |v_0\rangle \otimes |v_1\rangle \cdots \otimes |v_{q-1} \rangle  = |v_{q-1}\rangle \otimes |v_0\rangle \cdots \otimes |v_{q-2} \rangle.
\end{equation}
Two Hadamard gates $H$ and the controlled-$\Pi_q$ create a superposition of the permuted and non-permuted states, leading to the following output state
\begin{equation}
    \ket{\Psi_f} = \frac{1}{\sqrt{2}} \left( \ket{+} \otimes \ket{\phi_I} + \ket{-} \otimes \ket{\phi_C}\right),
\end{equation}
where $\ket{\pm} = \frac{1}{\sqrt{2}} (\ket{1}\pm \ket{0}) $,
\begin{equation}
\nonumber
     \ket{\phi_I} = \sum
     c_{i_{0}} \cdots c_{i_{q-1}} | i_0, i_1,\cdots, i_{q-1}\rangle \otimes | i_1,\cdots, i_{q-1}\rangle,
\end{equation} 
\begin{equation}
\nonumber
     \ket{\phi_C} =\sum
     c_{i_{0}} \cdots c_{i_{q-1}}  | i_{q-1}, i_0,\cdots, i_{q-2}\rangle \otimes | i_1,\cdots, i_{q-1}\rangle,
\end{equation}
and the sums in $\ket{\phi_{I,C}}$ extend over all the indices $i_0,\cdots,i_{q-1}$. Observing that $\langle \phi_C|\phi_C \rangle=1=\langle \phi_I|\phi_I \rangle$ due to the normalization of $\ket{\psi}$, while $\langle \phi_C|\phi_I \rangle = \sum_i |c_i|^{2q}$, we find the probability $P_0$ to find the ancilla qubit in state $|0\rangle$, given as $\langle \Psi_f| \left( | 0 \rangle \langle 0|  \otimes \mathbb{1}  \right)| \Psi_f \rangle$ reads 
\begin{equation}\label{Eq5}
   P_{0} = \dfrac{1}{2} + \dfrac{\sum_i | c_i |^{2q}}{2} = \frac{1}{2} + \frac{1}{2} I^{_Z}_q.
\end{equation}
 
Hence, the IPR $I^{Z}_q$ can be directly obtained from the single-qubit measurements. Repeating the measurement of the Pauli-Z operator on the ancillary qubit $N_{\text{s}}$ times, the average of the results approaches the value of $P_1$ up to a statistical uncertainty scaling as $\epsilon \sim N_{\text{s}}^{-1/2}$. To enhance accuracy and reduce statistical errors, one available approach is the implementation of quantum amplitude estimation~\cite{AmplitudeEstimation}. Our quantum algorithm can also be extended to an arbitrary basis $\{|b_i\rangle\}$. In that case, an additional unitary transform $V=\sum_{j} e^{-\text{i}\phi_j} |j\left\rangle \right\langle b_j|$ between computational basis and $\{|b_i\rangle\}$ is implemented before the $U_{\text{CNOT}}$ gates in Eq.\eqref{Eq4}. 
 
We note that for the specific choice of $q=2$, the controlled-$\Pi_2$ is equivalent to a controlled-SWAP gate. Then, the ancillary qubit implements the SWAP test protocol~\cite{swaptest}, returning fidelity between $|\psi\rangle$ and reduced quantum state $\rho = \sum_{i} |c_i|^{2} |i\rangle\langle i|$.  

Finally, we comment on the resources required by our algorithm. In comparison to the requirement of $L$ measurements required to estimate the probabilities $|c_i|^2 = | \langle i  \ket{\psi}|^2$ in the computational basis $\mathcal B_Z$, our algorithm requires only a single qubit measurement.
 
However, in general, the IPRs may be exponentially small in system size $L$. In such cases, exponentially small statistical error $\epsilon$, and consequently, an exponentially large $N_s$, may be required to achieve an accurate estimation of the IPRs. To illustrate this, we consider the following examples
\begin{itemize}
    \item a basis state $\ket{\sigma_i}$ in the computational basis, for which $I^{Z}_q=1$. In that case, statistical error $\sigma\approx O(1)$ is sufficient to reach a small relative error of $I^{Z}_q$;
    \item a GHZ state $\ket{\mathrm{GHZ}} = \frac{1}{\sqrt{2}}\left( |1\rangle^{\otimes L} +|0\rangle^{\otimes L} \right) $, for which $I^{Z}_q = 2^{1-q}$. In spite of the non-trivial entanglement structure of this state, a statistical error of $\sigma\approx O(1)$ is sufficient for any $q \geq 2$;
    \item a product state \begin{equation}
        \ket{\theta} = \left( \cos(\theta) \ket{0}+\sin(\theta)\ket{1}) \right)^{\otimes L};
    \end{equation}
    for which $I^{Z}_q = \left(\cos^{2q}(\theta)+\sin^{2q}(\theta) \right)^L$. For a generic value of $\theta$, the IPR is exponentially small in the system size $L$. To resolve such a quantity, an exponentially large  $N_s$ is required;
    \item random Haar state, for which $I^{Z}_q \propto 2^{(1-q)L}$. In that case,  $N_s$ scaling exponentially with $L$ is required for an accurate estimation of the IPRs.
\end{itemize}
These considerations show the practical difficulties encountered when estimating the IPRs, associated with the possible exponential smallness of the estimated quantity. This property reflects the fact that IPRs, by their construction, quantify the properties of many-body wave functions in the full many-body Hilbert space. As shown by the example of the product state, local rotations of the basis (e.g. fixing $\theta=0$) may dramatically decrease the resources needed for the estimation of IPRs. For certain tasks associated with assessing the dynamical properties of many-body systems, the eigenbasis of the Hamiltonian is distinguished, motivating the considerations of the following section.
 
\begin{figure}[t!]
	\centering
    \includegraphics[width=0.5\textwidth]{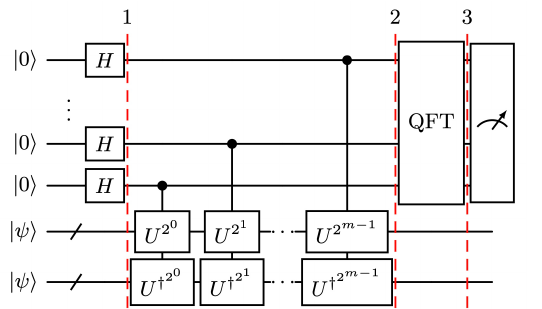} 
	\caption{
Quantum circuit for estimating the IPR $I^{\cal H}_2$ in the eigenbasis $\mathcal B_{\mathcal{H}}$ of Hamiltonian $\hat{\mathcal{H}}$. The quantum circuit operates on two copies of the considered state $\ket{\psi}$ and $m$ ancillary qubits, and involves the action of the evolution operator $U$ and QFT. Measurement of the ancillary qubits allows us to estimate the value of $I^{\cal H}_2$.
\label{fig:circuit2}
} %
	
\end{figure}

\subsection{IPR in eigenbasis of a selected Hamiltonian for qubits}\label{sec:IIIB}
In this section, we introduce a quantum algorithm for the calculation of IPR, denoted as $I^{\cal H}_2$ in an eigenbasis $\mathcal B_{\mathcal{H}}$ of a selected Hamiltonian $\hat{\mathcal{H}}$. When a many-body system is initially prepared in a state $\ket{\psi_0}$, a survival probability defined as $ |\langle 
\psi_0\ket{\psi(t)}|^2$, where $\ket{\psi(t)} = e^{-\text{i} \hat{\mathcal{H}} t}\ket{\psi_0}$, provides means of assessing ergodicity and ergodicity breaking in the system~\cite{Torres-Herrera15,Torres-Herrera18, Prelovsek18reduced, Tikhonov18stat, Schiulaz19, Santos20, Hopjan23survival, Pain23return, Creed23transportFS}. Using the eigendecomposition of the Hamiltonian, $\hat{\mathcal{H}} \ket{\varepsilon_i} = \varepsilon_i\ket{\varepsilon_i}$, and assuming the lack of degeneracy of the eigenvalues $\varepsilon_i $, we find that the long-time average of the survival probability, \begin{equation}\label{Eq0}
     \lim_{\tau \rightarrow \infty} \frac{1}{\tau} \int_{0}^{\tau} dt | \langle \psi (t)|  \psi_0 \rangle |^2 = \sum_{i} | \langle 
\varepsilon_i \ket{\psi_0}|^4= I^{{\cal H}}_2,
\end{equation}
is equal to the IPR with R\'{e}nyi index $q=2$ of the initial state $\ket{\psi_0}$ in the eigenbasis $\mathcal B_{\mathcal{H}} = \{ \ket{\varepsilon_i} \} $ of the Hamiltonian $\hat{\mathcal{H}}$. Notice that the Hamiltonian $\hat{\mathcal{H}}$ may be defined both for a system of qubits or for a system of qudits, since the part of the circuit—responsible for performing the Quantum Fourier Transform (QFT)—remains in qubit form regardless of the configuration on Hamiltonian. This motivates us to introduce the following quantum algorithm to calculate $I^{\cal H}_2$ in the basis $\mathcal B_{\mathcal{H}}$.


Without requiring prior knowledge of the eigenbasis $\mathcal{B}_{\mathcal{H}}$ of the Hamiltonian $\hat{\mathcal{H}}$, we estimate the second moment $I^{\mathcal{H}}_2$ using the quantum algorithm depicted in Figure~\ref{fig:circuit2}. The algorithm begins by preparing the initial state $|\Psi_0\rangle = |0\rangle^{\otimes m} \otimes |\psi\rangle^{\otimes 2}$, where $|0\rangle^{\otimes m}$ represents an $m$-qubit ancilla register and $|\psi\rangle^{\otimes 2}$ denotes two copies of the state whose properties we wish to characterize. Next, Hadamard gates are applied to all $m$ ancilla qubits, creating a uniform superposition over all possible computational basis states and effectively preparing the ancilla register to control subsequent operations. After the Hadamard layer, the circuit applies controlled-$U$ and controlled-$U^\dagger$ gates on the two copies of $|\psi\rangle$, where the unitary operator is defined as $U = e^{-i \hat{\mathcal{H}} t}$, and its power is conditioned on the binary string $x$ represented by the state of the ancilla qubits; for each ancilla basis state $|x\rangle$, with $x \in \{0, 1, \dots, 2^m - 1\}$, the operators $U^x \otimes (U^\dagger)^x$ are applied to the two copies of $|\psi\rangle$, thereby encoding phase information related to the eigenvalues of $\hat{\mathcal{H}}$ into the ancilla register. Finally, a Quantum Fourier Transform (QFT) is performed on the ancilla register, converting the phase information from the controlled operations into amplitude information that can be read out by a standard measurement in the computational basis; this transformation is critical for extracting the spectral properties necessary for estimating $I^{\mathcal{H}}_2$. In summary, the circuit in Figure~\ref{fig:circuit2} implements a variant of the quantum phase estimation algorithm, where the dashed red line indicates the intermediate cut showing the output state after the Hadamard gates that is used to control the time-evolution operations, and although the circuit is presented in a form familiar from the qubit setting, its structure naturally extends to the qudit case, as the underlying controlled operations and QFT remain analogous.

\begin{equation}
    |x\rangle \xrightarrow{\text{QFT}} \frac{1}{\sqrt{2^{m}}}\sum_{k=0}^{2^m -1 } e^{\text{i}\frac{2\pi x k}{2^m}} |k\rangle.
\end{equation}
At the end of the circuit, measurements on the product of Pauli-Z strings $\hat{Z}^{\otimes m}$ are implemented.

More concretely, the input state undergoes the following evolution (see Figure~\ref{fig:circuit2})
\begin{align}
    |\Psi_{0}\rangle \xrightarrow{1} &\dfrac{1}{\sqrt{2^{m}}} \sum_{x=0}^{2^m - 1} |x\rangle \otimes | \varphi \rangle \otimes | \varphi \rangle, \notag \\
    \xrightarrow{2} &\dfrac{1}{\sqrt{2^{m}}} \sum_{x=0}^{2^m - 1} |x\rangle \otimes U^{x} \sum_{i} c_i |\varepsilon_{i} \rangle \otimes U^{-x} \sum_{j} c_j |\varepsilon_{j}\rangle, \notag  \\ 
    = & \sum_{i, j} \dfrac{ c_ic_j }{\sqrt{2^{m}}}\sum_{x=0}^{2^m - 1} |x\rangle \otimes e^{-\text{i}\tilde{\varepsilon}_{ij} t x} |\varepsilon_{i} \rangle |\varepsilon_{j}\rangle, \notag \\ 
   \xrightarrow{3} &\sum_{i, j} \dfrac{c_ic_j}{2^{m}} \sum_{x=0}^{2^m - 1} \left( \sum_{k=0}^{2^m -1 } e^{\text{i}\frac{2\pi x k}{2^m}} |k\rangle \right)  \otimes e^{-\text{i} \tilde{\varepsilon}_{ij} x t} |\varepsilon_{i} \rangle |\varepsilon_{j}\rangle, \label{Eq6} 
\end{align}
where the energy difference  $(\varepsilon_i - \varepsilon_j)$ is abbreviated as $\tilde{\varepsilon}_{ij}$.
The probability to measure $\{0\}^{m}$ on the ancillary qubits on the output state $\Psi_{\text{out}}$ is given by
\begin{align}\label{Eq7}
   P_{{0, m}} &=  \left< \Psi_{\text{out}} \right| (\left| 0\right\rangle \left\langle 0\right|^{\otimes m} \otimes \mathbb{1})\left|\Psi_{\text{out}}\right\rangle 
   = I^{\cal H}_2 + \epsilon_r,
\end{align}
where $I^{{\cal H}}_2$ is the IPR of $\ket{\psi}$ in the eigenbasis $\mathcal{B}_{\mathcal H}$ and $0 \leq \epsilon_{r} \leq 4^{-m}\frac{\pi^2}{2\Delta^2t^2}$ with $\Delta$ denoting the minimum gap in the energy spectrum of $\hat{\mathcal{H}}$.

Notably, the upper bound of the error decreases exponentially with the number $m$ of ancillary qubits, which is a result of deploying an exponentially large in $m$ power of the operator $U$. The detailed proof of Eq.~\eqref{Eq7} and error analysis of $\epsilon_{r}$ are provided in Appendix~\ref{App:A}. The validity of the algorithm and its error bound in the presence of degeneracy will be discussed in the Appendix~\ref{App:C}.

The realization of the controlled-$U$ operation is a crucial step in the proposed quantum algorithm as it allows for encoding of the information about the eigenstates and the corresponding phase factors into the phases of ancilla qubit states.

The evolution operator $U$ may be implemented via Suzuki-Trotter  decomposition~\cite{Trotter59, Suzuki76}.
The involved approximation results in a discrepancy between the exact time-evolution operator $e^{-\text{i}\hat{\mathcal{H}}t}$ and its approximation $U_t$, which can be quantified via $\epsilon_{t} = \parallel U_t - e^{-\text{i}{\cal H} t} \parallel$. It can be shown that $\epsilon_{t} \leq \frac{\parallel {\cal H}  \parallel^2 t^2}{2 n_{T}}$ for the first-order Trotterization, with $n_{T}$ referring to the number of trotter steps~\cite{lloyd1996universal}. Assuming that the number of gates in first-order Trotterization is $N_{t}$, the total number of gates for the algorithm in Figure~\ref{fig:circuit2} is $\mathcal{O}(2^{m+1} N_{t} + m^2)$. 

\subsection{IPR in computational basis for qudits}\label{sec:IIIC}

\begin{figure}[t!]
	\centering
	\includegraphics[width=0.45\textwidth]{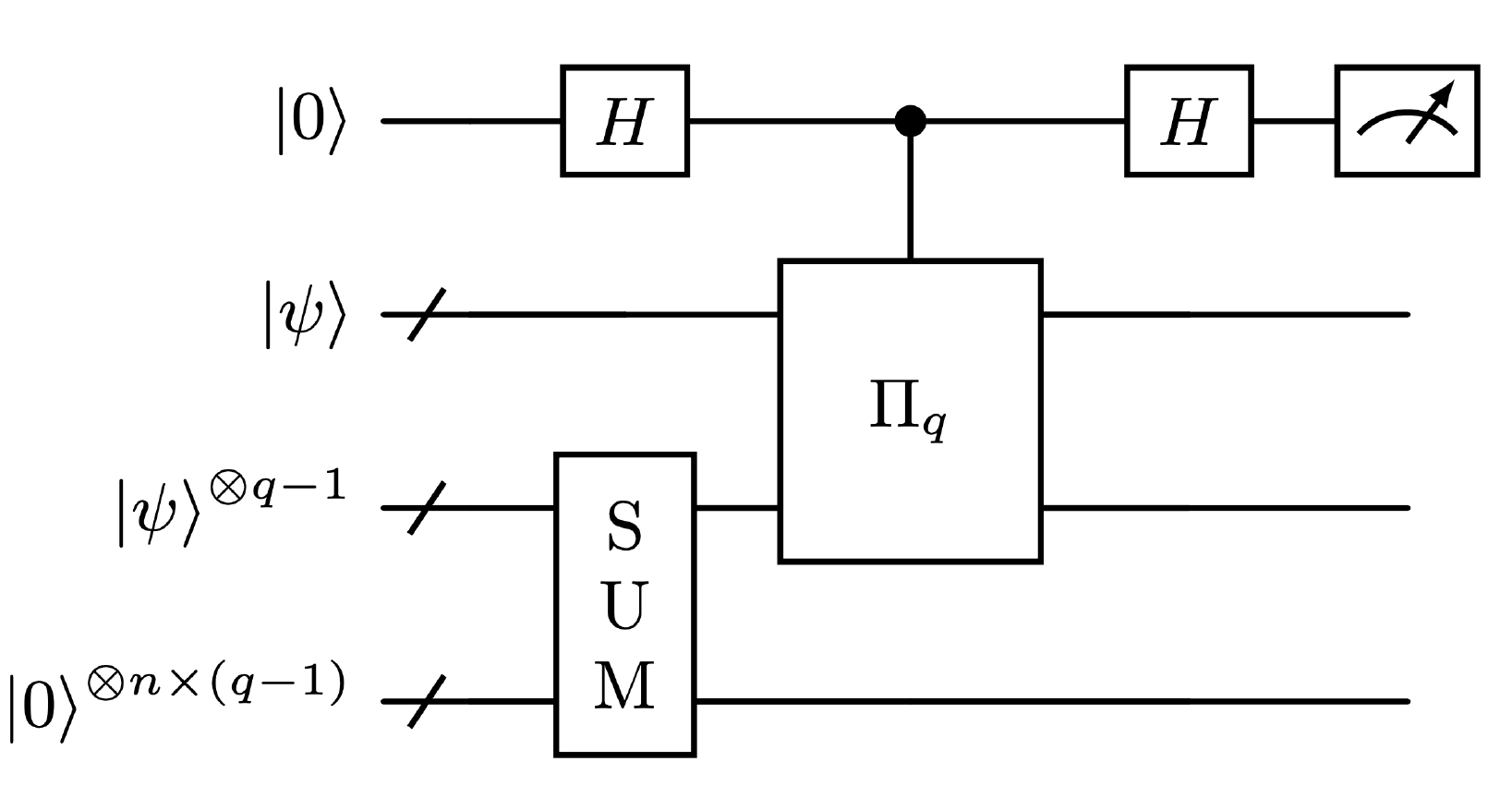} 
	\caption{Generalized quantum circuit for estimating IPR $I^Z_q$ in the computational basis $\mathcal B_Z$ on a $d-$dimensional qudit system. The quantum circuit comprises $n\times (q-1)$ $\text{SUM}_d$ gates and a controlled-$\Pi_q$ gate. The generalized quantum circuit kept the first system as a qubit, while the rest of the line represents qudit systems. The value of $I^Z_q$ is determined as an expectation value of the Pauli-Z operator on the first qubit.}
	\label{fig:circuit3}
\end{figure}

Qudit-based quantum simulators, particularly those utilizing trapped ions \cite{Hrmo2023} and superconducting circuits \cite{Goss2022}, hold great promise for achieving quantum advantage in the near future \cite{Ringbauer2022}. Qudits also offer a valuable approach for simulating physical systems that are not inherently formulated in a qubit basis. Although every qudit system can theoretically be mapped to qubit systems, the associated technical complexities may make their simulation impractical. For instance, lattice gauge theory models are naturally expressed in qudit language once the continuous variable truncation is executed \cite{Funcke_2021, dimeglio2023quantum, PRXQuantum_4_027001}.

Extension of our algorithms to the case of qudits with local Hilbert space dimension $d>2$ is possible.
In the following, we generalize our algorithm in Figure~\ref{fig:circuit3} to estimate $I^{Z}_q$ in the computational basis of qudit systems.

We denote the computational basis of a $d-$dimensional qudit system as $\mathcal{B}_{Z} = \{\ket{\sigma_i}\}$, where $\sigma_i=0,1,2, \cdots, d-1$. Then, the CNOT gate can be generalized to $\text{SUM}_d$ gate~\cite{wang2020qudits}
\begin{equation}
    \text{SUM}_d \ket{\sigma_i} \otimes \ket{\sigma_j} = \ket{\sigma_i} \otimes \ket{\sigma_i + \sigma_j (\bmod \ d)}.
\end{equation}
To measure the value of $I^{Z}_q$, we replace the CNOT gates in Figure~\ref{fig:circuit1} by the $\text{SUM}_d$ gates for qudit systems, while the ancillary system, i.e. the top qubit in Figure~\ref{fig:circuit1} is still kept as a binary system. Finally, the two Hadamard gates in the first line remain intact as well.
Application of the $\text{SUM}_d$ gate on the  state $\ket{\sigma_j}\otimes\ket{0}$ yields
\begin{align}
    \text{SUM}_d \ket{\sigma_j}\otimes\ket{0} = \ket{\sigma_j}\otimes\ket{\sigma_j}.
\end{align}
This allows us to generate the entanglement between the $(q-1)$ copies of $|\psi\rangle$ and the $n$-qudit quantum registers by the $\text{SUM}_d$ gates,
\begin{align}
    \text{SUM}^{\otimes n (q-1)}_d |\Psi_{0}\rangle =&\sum_{i_0,\cdots, i_{q-1}} c_{i_{0}} \cdots c_{i_{q-1}} |0\rangle \otimes | i_0 \rangle \notag\\
     &\otimes | i_1,\cdots, i_{q-1}\rangle \otimes | i_1,\cdots, i_{q-1}\rangle,
\end{align}
in accordance with the Eq.~\eqref{Eq4}. The rest
of the quantum algorithm follows closely the qubit case, leading to the same result 
as in Eq.~\eqref{Eq5}. The complexity analysis for the qudit case follows similar principles to the qubit case. The required number of measurements \(N_s\) remains the same as stated in the previous section: \(N_s \in \mathcal{O}\left(1/\epsilon^2\right)\), since the qudit case also relies solely on single-qubit measurements. We also note that the sample complexity (i.e. required number of original copies of $| \psi \rangle$) for estimating $I^{Z}_q$ scales as $\mathcal{O}\left(q/\epsilon^2 \right)$ since each run of the quantum circuit takes $q$ copies of $| \psi \rangle$. It is seen that the sample complexity is independent of the system size, but only constantly scales with the desired order of the Rényi index and inversely scales with the square of the error tolerance $\epsilon$.

\section{Applications and Examples}\label{sec:IV}
In this section, we give numerical results demonstrating the applicability of the algorithms developed in Sec.\ref{sec:III},
i.e. we compare exact diagonalization (ED) results with the quantum algorithms simulations. The first two examples are for qubit systems, i.e. the one-axis twisting dynamics in spin-$1/2$ systems, Sec.\ref{sec:IVA}, and the ergodicity in the extended PXP model, Sec.\ref{sec:IVB}. In the last subsection Sec.\ref{sec:IVC} we consider qudit systems and probe the ground state of the spin-$1$ AKLT model. 

\subsection{Simulate one-axis twisting dynamics}\label{sec:IVA}
 
In the following, we employ the quantum algorithm in Figure~\ref{fig:circuit1} to simulate one-axis twisting (OAT) in spin systems ~\cite{kitagawa1993squeezed,wineland1994squeezed}. OAT has been studied extensively in theory and experiments, showing its applications in quantum information science and high-precision metrology~\cite{pezze2018quantum,wolfgramm2010squeezed,wineland1994squeezed,muller2023certifying}. The OAT protocol dynamically generates non-trivial quantum correlations via time evolution
    $|\psi(t)\rangle =  e^{-\text{i} t\hat{\mathcal{H}}_{\rm OAT}}(t)|\psi_0\rangle$,
where 
$\hat{\mathcal{H}}_{\rm{OAT}} =   \frac{1}{4}\sum_{i,j=1}^L \hat{Z}_i \hat{Z}_j$, with the initial spin coherent state, i.e. state with all spins polarized along $x$-direction, $|\psi_0\rangle=|0\rangle^{\otimes L}_x$. The OAT protocol generates spin-squeezed, many-body entangled, and many-body Bell-correlated states~\cite{schmied2016bell, baccari2019bell, tura2014detecting, aloy2019device, muller2021inferring, niezgoda2020quantum, niezgoda2021many, plodzien2020producing, plodzien2022one, yanes2022one,  PhysRevA.107.013311, PhysRevB.108.104301, yanes2024exploring,plodzien2024generation,PhysRevA.111.012417}. In particular,  at time $t=\pi/4$, the generation of the L-body GHZ (Greenberg-Horne-Zeilinger) state is created  along $x$-direction, $\ket{\mathrm{GHZ}} = \frac{1}{\sqrt{2}}\left( |1\rangle^{\otimes L}_x +|0\rangle^{\otimes L}_x \right) $.
The system's dynamics can be investigated  by  measuring the IPR $I^X_2$ 
in the eigenbasis ${\cal B}_X$ of Pauli-X $\hat{X}_i$ operators, obtained by a local rotation of the computational basis $\mathcal B_Z$. The OAT evolution  interpolates between the ${\cal B}_X$ basis spin coherent state $|1\rangle^{\otimes N}_x$, for which $I^{X}_2=1$, and the $\ket{{\rm GHZ}}$ state  for which the  $I^{X}_2$ admits value $\frac{1}{2}$.

\begin{figure}
	\centering 
	\includegraphics[width=0.45\textwidth]{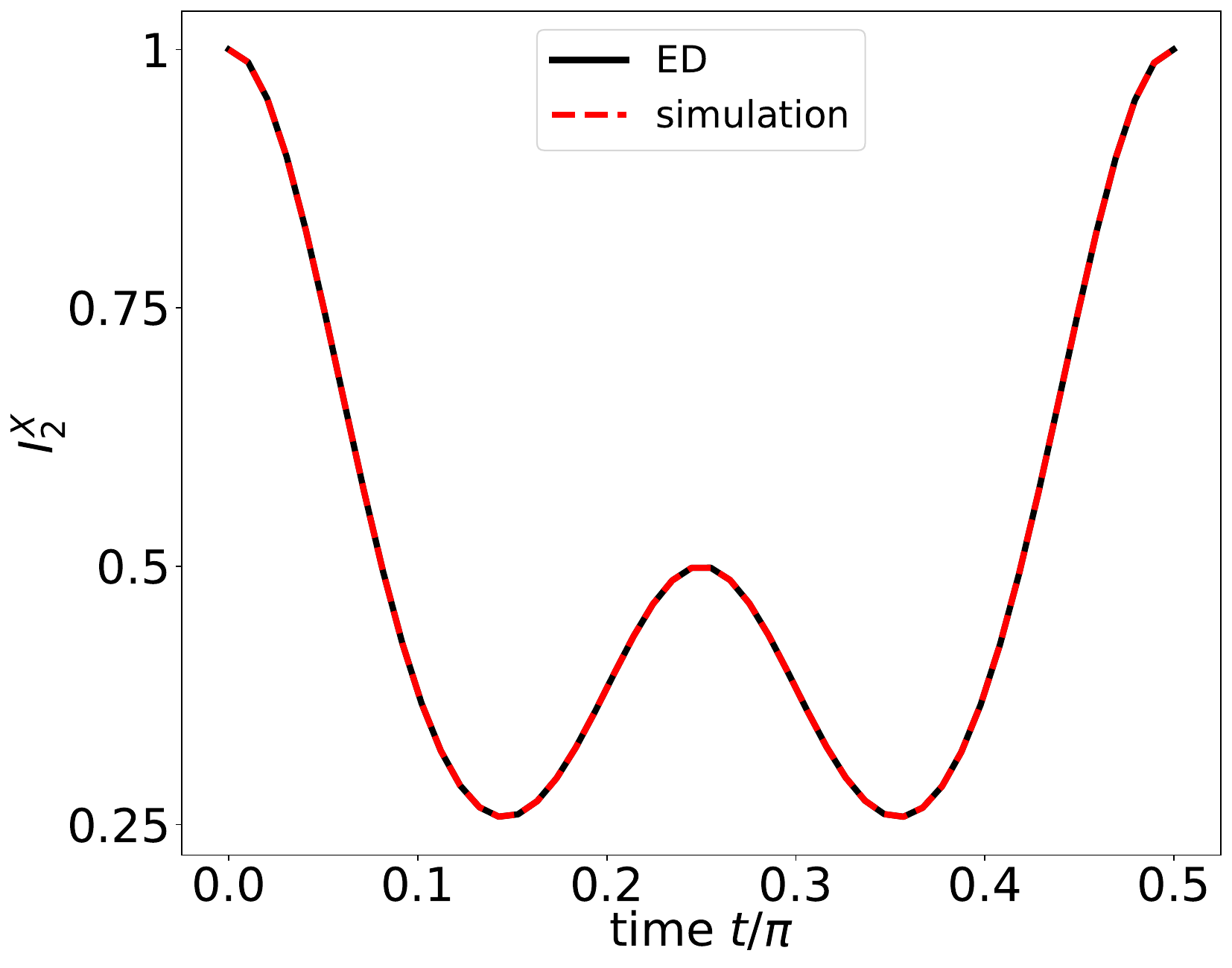} 
	\caption{Time-evolution of the IPR $I^X_2$ during the one-axis twisting protocol  for $L=4$ spins-$1/2$. At time $t=0$ the system is prepared with all spins polarized along $x$-direction having IPR is $I^X_2=1$. At $t=\frac{\pi}{4}$ the GHZ state along $x$-direction is generated, and $I^X_2 = \frac{1}{2}$. The black-solid line ED represents the exact diagonalization results, while the red dashed line corresponds to the results obtained from simulating the quantum algorithm presented in Figure~\ref{fig:circuit1}.
 }
	\label{fig4} 
\end{figure}

In Figure~\ref{fig4}, we present the OAT dynamics for the case of $L=4$ qubits. In accordance with Eq.~\eqref{Eq5}, the numerical simulation of the quantum algorithm of Figure~\ref{fig:circuit1} accurately reproduces the ED curve.

\subsection{Probing ergodicity in the extended PXP model}\label{sec:IVB}
To exemplify the practical utilities of the proposed algorithms, we first implement the algorithm in Figure~\ref{fig:circuit2} to investigate the 
thermalization in a PXP model with Zeeman magnetic field,  
with Hamiltonian given as:
\begin{equation} \label{eq:PXP}
	 \hat{\mathcal{H}} = \sum_{i=1}^{L} \left( \hat{P}_{i-1} \hat{X}_{i} \hat{P}_{i+1} - h  \hat{Z}_{i} \right),
\end{equation}
where $\hat{X}_i,\hat{Z}_i$ denotes Pauli-X,Z operators acting on the $i$-th spin, $\hat{P}_i $ is the projector on the $\ket{0}$ state of the $i$-th spin, 
and $h \in [0, 1]$ is the amplitude of the external transverse field. We assume periodic boundary conditions.
In the absence of the external field, for $h=0$, the PXP model is known as a paradigmatic model of quantum many-body scars~\cite{Serbyn21}. The presence of the scar states is manifested as a lack of thermalization when the system is initialized in particular states, for instance in the N\'{e}el state~$ | 0101 \cdots \rangle$~\cite{Bernien2017}. In contrast, for generic initial conditions, the system thermalizes similarly to other interacting non-integrable many-body systems~\cite{Rigol08}. The quantum many-body scars states form a ladder of highly excited eigenstates extending over the whole spectrum of $\hat{\mathcal{H}}$~\cite{Turner18}. The ground state of the model Eq.\eqref{eq:PXP} undergoes a quantum phase transition of Ising universality class at $h_{c} \approx 0.655$~\cite{criticality2022}. 

The properties of the system in the vicinity of the quantum phase transition in the ground states are linked with the behavior of the quantum many-body scars in~\cite{criticality2022}. The thermalization of the N\'{e}el state under the time evolution generated by Eq.\eqref{eq:PXP} was probed with $\delta \sigma^{z}$ representing the difference between the long-term average of the operator $\hat{Z}$ and the thermal equilibrium expectation value $Z_{\rm th}$:
\begin{equation}
    \delta \sigma^{z}= \langle \hat{Z}\rangle-Z_{\rm th}.
\end{equation}
The behavior of $\delta \sigma^{z}$ at fixed system size $L$ may be summarized as follows~\cite{criticality2022}: below the transition, at $h<h_c$, lack of thermalization due to the presence of scar states is observed, at the criticality $h \approx h_c$ the system thermalizes, while for  $h>h_c$ lack of thermalization of the system occurs due to a high overlap of the N\'{e}el state with the ground state of the system.
 
\begin{figure}[t!]
	\centering
 \includegraphics[width=0.99\linewidth]{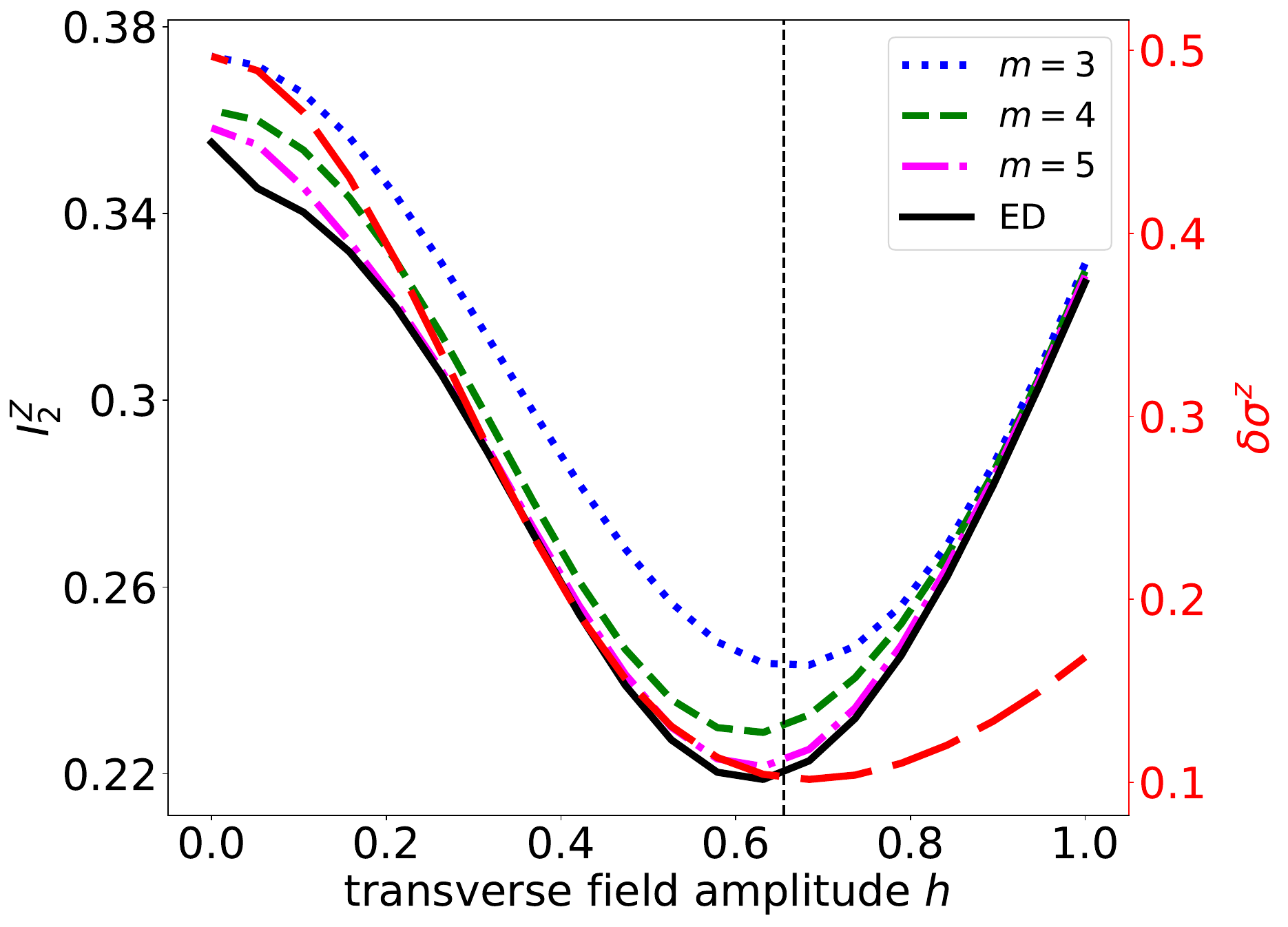} 
	\caption{
 Thermalization and ergodicity breaking in PXP model in the presence of a transverse field  $h$, Eq.\eqref{eq:PXP}. Left $y$-axis: The IPR $I^{\cal H}_2$ in the eigenbasis $\mathcal B_{\mathcal H}$ of the Hamiltonian is calculated with the algorithm of Figure~\ref{fig:circuit2}. Simulations are conducted for system size $L=8$, with a fixed $t=1$, and with the number of Trotter steps $N_t = 10$. 
 The constituent lines (from top to bottom): dotted-blue, dashed-green, dash-dotted-pink, correspond to outcomes obtained by estimation with $m =3, 4, 5$ ancillary qubits, respectively.
 The black solid line corresponds to ED results.
Right red color $y$-axis: long-dashed-red line presents the difference $\delta\sigma^z$ between the long-time average and the thermal value of the Pauli-Z operator as a function of the transverse field strength for $L = 8$ spins; vertical line corresponds to critical $h_c \approx 0.655$ ~\cite{criticality2022}.}
	\label{fig5} 
\end{figure}
 
The thermalization of the system may be also probed by the value of IPR $I^{{\cal H}}_2$ of the initial state $|\psi_0\rangle = | 0101 \cdots \rangle $ in the eigenbasis $\mathcal B_{\mathcal H}$ of the PXP Hamiltonian Eq.~\eqref{eq:PXP}, equal to the long-time average of the survival probability $ |\langle \psi_0\ket{\psi(t)}|^2$. The results of the algorithm of Figure~\ref{fig:circuit2} are presented in Figure~\ref{fig5}. We fix  $t=1$ and set the number of trotter steps as $N_t = 10$, and compare the obtained value of $I^{{\cal H}}_2$ with the exact value of IPR calculated with the exact diagonalization of $\hat{\mathcal{H}}$. As it is expected from Eq.~\eqref{Eq7}, with the increasing number $m = 3, 4, 5$ of the employed ancillary qubits, the results approach the exact diagonalization value~\cite{Github}. Moreover, the value of IPR decreases monotonically with $h$ in the whole interval $h \lesssim h_c$, showing that the thermalization of the system is more effective as the critical regime $h \approx h_c$ is approached. At $h \approx h_c$, the IPR $I^{{\cal H}}_2$ admits a minimal value, and increase at larger $h$, consistently with the behavior of $\delta \sigma^{z}$. 

These results show that the proposed algorithm can be useful in probing thermalization and ergodicity breaking in quantum many-body systems. 

\subsection{IPR for AKLT model}\label{sec:IVC}

Here, we present an algorithm for obtaining $I_2^Z$ for the qudit system, with on-site Hilbert space dimension $d=3$. We consider   
the spin-$1$ chain described by the
AKLT   (Affleck-Kennedy-Lieb-Tasaki) model \cite{PhysRevLett.59.799,Affleck1988} with the transverse field and with the open boundary conditions:
\begin{equation}\label{eq:H_AKLT}
    \hat{\cal H}\!=\! \sum_{i=1}^{L-1}
\left[\frac{1}{2}\mathbf{\hat{S}}_i\cdot\mathbf{\hat{S}}_{i+1}\! + \!\frac{1}{6} \big( \mathbf{\hat{S}}_i\cdot\mathbf{\hat{S}}_{i+1}\big)^2 \!+ \!\frac{1}{3} \right] -\dfrac{h}{L}\sum^{L}_{i=1} \hat{S}_i^z.
\end{equation}
For $h=0$, the ground state of the Hamiltonian is a valence bond solid where each neighboring site pair is linked by a single valence bond. With open boundary conditions, the edge spins-$1$ have only one neighbor, leaving one of their constituent spin-$1/2$ unpaired (for review see \cite{Wei2022}).

\begin{figure}
	\centering
 \includegraphics[width=0.99\linewidth]{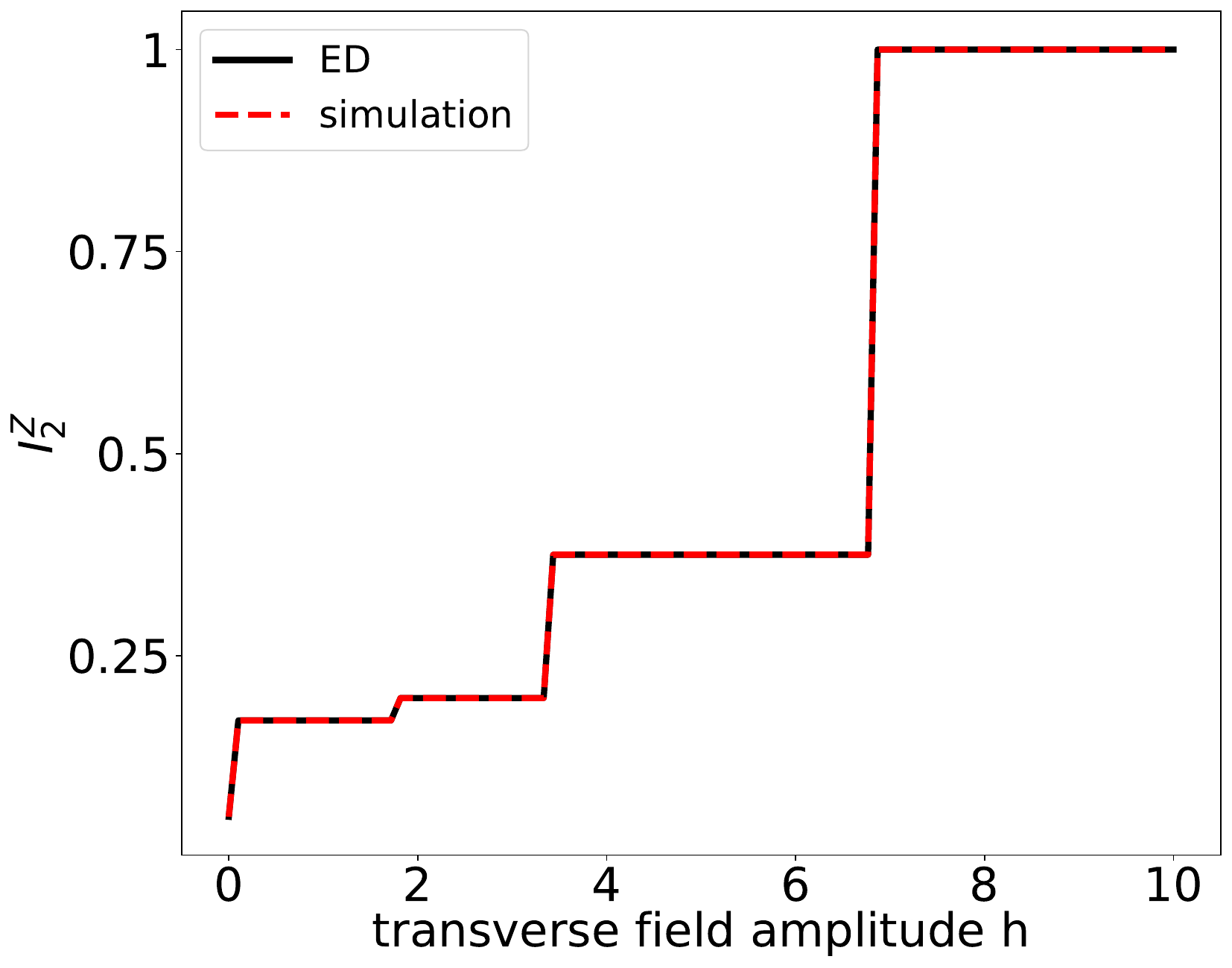} 
	\caption{
  The IPR $I^Z_2$ of the AKLT ground state, Eq.\eqref{eq:H_AKLT}, for $L=4$ spins-$1$, as a function of the transverse field $h$.
 The black solid line corresponds to ED results, while the red dashed line corresponds to the results of the quantum algorithm in Figure~\ref{fig:circuit3}. 
 }
	\label{fig:fig6} 
\end{figure}

The AKLT state is a paradigmatic example of a symmetry-protected topological (SPT) order \cite{PhysRevB.85.075125}. 
AKLT states play a role in a measurement-based quantum computation \cite{PhysRevLett.86.5188, PhysRevLett.98.220503, PhysRevLett.105.020502}, where the computation begins in an appropriately entangled state, such as a g ground state of quantum spin chains with symmetry-protected topological order \cite{PhysRevLett.119.010504, MIYAKE20111656, PhysRevA.90.042333}, followed by a set of proper single particle measurements.
Recently, it has been shown that the AKLT ground state can be effectively prepared on a quantum circuit \cite{PRXQuantum.4.020315}.

In Figure~\ref{fig:fig6} we present the estimation of $I_2^Z$ for the ground state of the AKLT model as a function of the transverse field $h$ for a chain of $L=4$ spins-$1$. The results obtained via ED are in perfect agreement with the proposed algorithm, Figure~\ref{fig:circuit3}. The ground state is perfectly localized $I^Z_2 \to 1$ on a single state of $\mathcal{B}_Z$ for $h \gg 1$, and spreads over an increasing number of states $I^Z_2 < 1$  of $\mathcal{B}_Z$ for smaller values of the transverse field.

\section{Conclusion}
\label{sec:V}
In this work, we have introduced three quantum algorithms to estimate IPRs and participation entropies of a state of multi-qubit and multi-qudit system.  
We first focused on a case of a fixed known basis, such the as computational 
allowing for the estimation of IPRs 
on quantum devices. 
The introduced algorithms enable the estimation of IPRs with just single-qubit measurements. We exemplified the utility of the introduced algorithms for non-equilibrium quantum many-body problems by investigating the OAT dynamics.
Motivated by the relation of the IPR in the eigenbasis of a given Hamiltonian to the long-time average of survival probability, we introduced a quantum algorithm allowing estimation of the IPR in the Hamiltonian's eigenbasis without the necessity of the full diagonalization of the Hamiltonian.
We have shown that the estimation error diminishes exponentially with the addition of ancillary qubits paralleled by the deployment of high powers of the evolution operator. 
To validate the efficacy of this approach, we conducted simulations of a deformed PXP model. As the number of ancillary qubits increases, the estimated IPR closely aligns with exact diagonalization results, effectively capturing the system's thermalization properties. 
Finally, we presented the utilization of our algorithm for multiqudit systems analyzing the ground state of the spin-$1$ AKLT model in the transverse field, showing the perfect agreement between exact results and the proposed algorithm.

In future research, exploring applications of our algorithms in diverse areas such as quantum chemistry, condensed matter physics, and quantum computing optimization tasks could uncover new insights and further validate the effectiveness of quantum simulations in the NISQ era. 

\section{Acknowledgement}
We acknowledge the beneficial discussion with Zhixin Song and Xhek Turkeshi. 

ICFO group acknowledges support from: European Research Council AdG NOQIA; MCIN/AEI (PGC2018-0910.13039/501100011033, CEX2019-000910-S/10.13039/501100011033, Plan National FIDEUA PID2019-106901GB-I00, Plan National STAMEENA PID2022-139099NB, I00, project funded by MCIN/AEI/10.13039/501100011033 and by the “European Union NextGenerationEU/PRTR" (PRTR-C17.I1), FPI); QUANTERA DYNAMITE PCI2022-132919, QuantERA II Programme co-funded by European Union’s Horizon 2020 program under Grant Agreement No 101017733; Ministry for Digital Transformation and of Civil Service of the Spanish Government through the QUANTUM ENIA project call - Quantum Spain project, and by the European Union through the Recovery, Transformation and Resilience Plan - NextGenerationEU within the framework of the Digital Spain 2026 Agenda; Fundació Cellex; Fundació Mir-Puig; Generalitat de Catalunya (European Social Fund FEDER and CERCA program; Barcelona Supercomputing Center MareNostrum (FI-2023-3-0024); Funded by the European Union. Views and opinions expressed are however those of the author(s) only and do not necessarily reflect those of the European Union, European Commission, European Climate, Infrastructure and Environment Executive Agency (CINEA), or any other granting authority. Neither the European Union nor any granting authority can be held responsible for them (HORIZON-CL4-2022-QUANTUM-02-SGA PASQuanS2.1, 101113690, EU Horizon 2020 FET-OPEN OPTOlogic, Grant No 899794, QU-ATTO, 101168628), EU Horizon Europe Program (This project has received funding from the European Union’s Horizon Europe research and innovation program under grant agreement No 101080086 NeQSTGrant Agreement 101080086 — NeQST); ICFO Internal “QuantumGaudi” project;

\appendix 
\newpage

\section{Error analysis for $\varepsilon_r$ in Sec.\ref{sec:IIIB}}\label{App:A}
Here we will prove that the error $\epsilon_{r}$ is bounded by $ 0 \leq \epsilon_{r} \leq 4^{-m}\frac{W^2}{\Delta^2}$.  The probability $P_{0, m }$ takes:
\begin{equation*}
\begin{aligned}
 P_{0, {m}} &=  \left< \Psi_{\text{out}}| (|0\right\rangle \left\langle 0|^{\otimes m} \otimes \mathbb{1})|\Psi_{\text{out}}\right\rangle \notag\\
 &=  \sum_{i, i^{'},j, j^{'}} \frac{c_ic_jc^{*}_{i'}c^{*}_{j'}}{4^{m}}  \sum_{x, x'=0}^{2^m - 1} e^{\text{i}(\tilde{\varepsilon}_{ij}x -\tilde{\varepsilon}_{i'j'} x' )t}  \langle \varepsilon_{j'} |\varepsilon_{j} \rangle \langle \varepsilon_{i'}  |\varepsilon_{i}\rangle \notag\\
&= \dfrac{1}{4^{m}} \sum_{i,j} |c_i|^2|c_j|^2 \sum_{x, x'=0}^{2^m - 1} e^{\text{i}\tilde{\varepsilon}_{ij}(x -x')t} \notag\\
&= \sum_{i} |c_i|^4 + \dfrac{1}{4^{m}} \sum_{i\neq j} |c_i|^2|c_j|^2 \sum_{x, x'=0}^{2^m - 1} e^{\text{i}\tilde{\varepsilon}_{ij}(x -x')t},
\end{aligned}
\end{equation*}
here the first term is $I^{\mathcal{H}}_2$, and the second term is  $\epsilon_r$ in Eq.~\eqref{Eq7}. Further simplification leads to:

\begin{equation*}
\begin{split}
    \epsilon_{r} &=  \dfrac{1}{4^{m}} \sum_{i\neq j} |c_i|^2|c_j|^2 \sum_{x, x'=0}^{2^m - 1} e^{\text{i}\tilde{\varepsilon}_{ij}(x -x')t} \notag\\
    &= \dfrac{1}{4^{m}} \sum_{i\neq j} |c_i|^2|c_j|^2 \sum_{x=0}^{2^m - 1} e^{\text{i}\tilde{\varepsilon}_{ij}xt} \sum_{x'=0}^{2^m - 1} e^{\text{i}\tilde{\varepsilon}_{ij}x't}\notag\\ 
    &=  \dfrac{1}{4^{m}} \sum_{i\neq j} |c_i|^2|c_j|^2 \big( \dfrac{1-e^{\text{i}\tilde{\varepsilon}_{ij} 2^{m}t}}{1-e^{\text{i}\tilde{\varepsilon}_{ij}t}} \big)\big( \dfrac{1-e^{-\text{i}\tilde{\varepsilon}_{ij} 2^{m}t}}{1-e^{-\text{i}\tilde{\varepsilon}_{ij}t}} \big)   \notag\\
    &= \dfrac{1}{4^m} \sum_{i\neq j} |c_i|^2|c_j|^2 \dfrac{1 - \cos{(2^m \tilde{\varepsilon}_{ij}t})}{1 - \cos{( \tilde{\varepsilon}_{ij}t )}}.  
\end{split}
\end{equation*}
For any real $\theta$, $1-\cos\theta \leq 2$, so
    $\epsilon_{r} \leq \dfrac{1}{4^m} \sum_{i \neq j} |c_i|^2|c_j|^2 \dfrac{2}{1 - \cos{( \tilde{\varepsilon}_{ij}t )}}$.  
The inequality above can be further bounded when $|\tilde{\varepsilon}_{ij}t|
\in (0, \pi], \forall i\neq j $ since $1-\cos\theta \geq \frac{2\theta^2}{\pi^2}$ for $\theta \in [-\pi, \pi]$:
\begin{equation*}
\begin{aligned}
     \epsilon_{r} &\leq \dfrac{1}{4^m} \sum_{ij} |c_i|^2|c_j|^2  \dfrac{\pi^2}{\tilde{\varepsilon}_{ij}^2 t^2} \notag \\ 
     &\leq \dfrac{1}{4^m} \sum_{ij} |c_i|^2|c_j|^2  \dfrac{\pi^2}{(\varepsilon_{i} - \varepsilon_{j})^2 t^2} \notag \\
    & \leq \dfrac{1}{4^m} \sum_{ij} |c_i|^2|c_j|^2  \dfrac{\pi^2}{\Delta^2 t^2}   \\
    & \leq 4^{-m}  \dfrac{\pi^2}{\Delta^2 t^2}.
\end{aligned}
\end{equation*}
Apparently, $\epsilon_{r} \geq 0$, then we finish the proof of $ 0 \leq \epsilon_{r} \leq 4^{-m}\frac{\pi^2}{\Delta^2 t^2}$.

\section{Degenerate case}\label{App:C} 

The algorithm in Hamiltonian basis and related error analysis are also valid in the presence of degeneracy. The Inverse Participation Ratio (IPR), as defined in our work, is expressed as:

\begin{equation}
I^{A}_q = \sum_{i=0}^{\mathcal{N} - 1} |c_i|^{2q}, \quad q = 2, 3, 4, \dots
\end{equation}

For a degenerate Hamiltonian $\mathcal{H}$, the spectral decomposition is given by:

\begin{equation}
\mathcal{H} = \sum_j \varepsilon_j \sum_{\alpha = 1}^{d_j} 
|\varepsilon^{\alpha}_j \rangle \langle \varepsilon^{\alpha}_j | 
= \sum_j \varepsilon_j P_j,
\end{equation}
where $d_j$ denotes the degeneracy degree of the $j$-th subspace, and $P_j$ is the projector onto the corresponding degenerate eigenspace. For a quantum state 
$|\psi\rangle = \sum_j \sum_{\alpha = 1}^{d_j} c_{j, \alpha} |\varepsilon^{\alpha}_j \rangle$, IPR can be generalized using these projectors:

\begin{equation}
I^{\mathcal{H}}_q = \sum_j |\langle \psi | P_j | \psi \rangle|^{2q}, \quad q = 2, 3, 4, \dots,
\end{equation}
which naturally extends the IPR definition to the degenerate case while remaining consistent with the non-degenerate scenario. 

Our algorithm estimates the IPR in terms of degenerate eigenspaces by computing:

\begin{equation}
P_{0, m} = \sum_j \left( \sum_{\alpha} |c_{j, \alpha}|^2 \right)^2 + \varepsilon_r 
= \sum_j |\langle \psi | P_j | \psi \rangle|^2 + \epsilon_r.
\end{equation}
Similar to the analysis in Appendix~\ref{App:A}, the error term $\epsilon_r$ is also bounded by 
\begin{equation}
0 \leq \epsilon_r \leq \frac{\pi}{4^m \Delta^2 t^2},
\end{equation}
while $\Delta$ represents the minimum gap between two non-degenerate energy levels. This formulation allows our algorithm to accurately estimate the IPR even in degenerate eigenspaces. When the quantum state resides entirely within a degenerate eigenspace, the IPR reaches its maximum value $I^{\mathcal{H}}_2 = 1$, confirming the algorithm's reliability in detecting localization in degenerate cases.

Moreover, our algorithm is capable of detecting the presence or absence of degeneracy in a varying Hamiltonian. This is achieved because the precision of the algorithm is sensitive to the minimum nonzero energy gap, particularly during transitions between degenerate and non-degenerate regimes. This sensitivity underscores the utility of our approach in probing and characterizing quantum systems with varying symmetries and energy spectra.

\section{OAT experiment on IBM quantum machine}\label{App:B}
\begin{figure}[!t]
	\centering
 \includegraphics[width=0.8\linewidth]{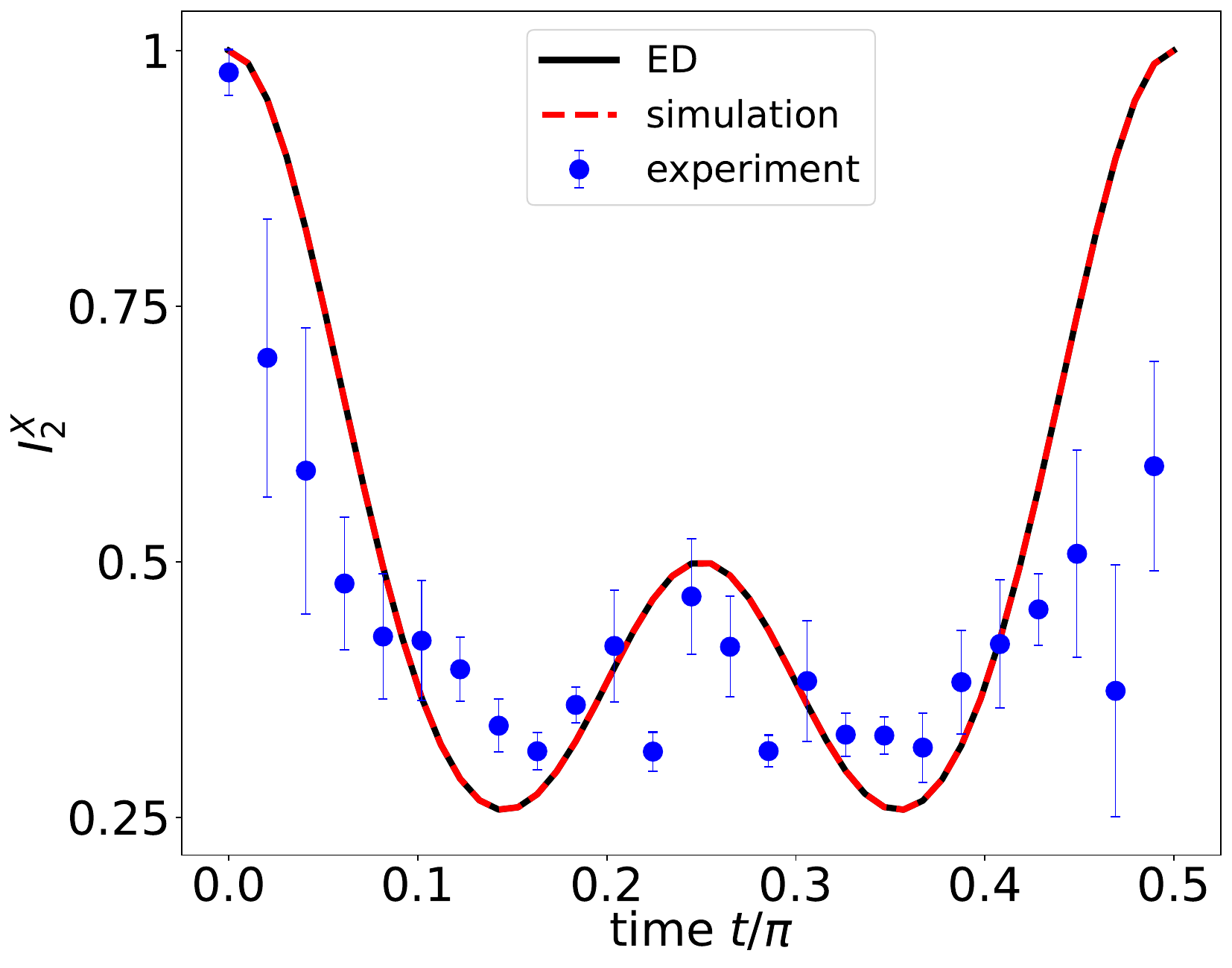} 
	\caption{
 Time evolution of the IPR $I^X_2$ for the OAT model, Sec.\ref{sec:IVA}.
The blue markers with error bars present the experiments run on ibm$\_$torino backend at IBM quantum platform, \cite{IBM},  by executing the quantum algorithm Figure~\ref{fig:circuit1} after averaging over $5$ runs at each time-stamp.
 }
	\label{fig:fig7} 
\end{figure}
Here, we present an experimental realization of a quantum algorithm, as shown in Figure~ \ref{fig:circuit1}, conducted on an IBM quantum machine. This involves measuring the IPR $I^X_z$ for the one-axis twisting protocol, detailed in Section \ref{sec:IVA}.

In Figure~\ref{fig:fig7}, the blue markers with error bars represent the estimated values of $I^{X}_2$. These estimations qualitatively align with the exact diagonalization (ED) results, depicted by the black solid line, and with the numerical simulations of the algorithm, indicated by the red dashed line. The experiment was conducted on IBM's ibm\_torino platform, with each data point derived from 2048 measurement shots per time step. We implemented noise mitigation strategies, including dynamical decoupling, and optimized the circuit transpilation as described in \cite{IBM}.

Transpiling the quantum algorithm from Figure \ref{fig:circuit1} into native quantum gates notably increases the circuit depth. The individual outliers observed in Figure \ref{fig:fig7} are linked to instances where transpilation at certain time splits ($t$) produced circuits significantly deeper than others, thereby heightening their susceptibility to noise. At $t=0$, where the evolution is not executed, the overhead from gates is equivalent to that of an idle gate, which results in a more precise measurement outcome. For further details on the numerical implementation, please refer to the GitHub repository cited in \cite{Github}.

The experimentally obtained value of $I^{X}_2 \approx \frac{1}{2}$ at $t \approx \pi/4$ substantiates the efficacy of the introduced quantum algorithm for exploring many-body dynamics on future fault-tolerant quantum computers.

\newpage
\nocite{*}
\bibliography{bibliography}
\end{document}